\documentstyle[epsfig]{aipproc}
\begin{document}
\title{Results from pQCD for A+A\\ collisions at RHIC \&
LHC\\ energies}

\author{K. Tuominen\footnote{kimmo.tuominen@phys.jyu.fi}}
\address{Department of Physics, University of Jyv\"askyl\"a,\\
	 P.O. Box 35, FIN-40351, Jyv\"askyl\"a, Finland}

\maketitle

\begin{abstract}
This talk will discuss how to compute initial quantites in heavy ion
collisions at RHIC (200 AGeV) and at LHC (5500 AGeV) using
perturbative QCD (pQCD) by including the next-to-leading order (NLO)
corrections and a dynamical determination of the dominant physical
scale. The initial numbers are converted into final ones by assuming
kinetic thermalization and adiabatic expansion.
\end{abstract}

The whole heavy ion physics community has entered into exciting and
intense period as the first results from RHIC are expected to appear,
and the era of LHC seems no more very distant. However, despite the
solid status of QCD as the theory of strongly interacting matter, many
uncertainties remain in the predictions for the outcome of the current
experiments and various methods have been applied. Further
experimental data will (hopefully) single out the best candidates for
the correct approach.

As the collision energy is increased from that of the SPS, larger
intrinsic scales are generated and the applicability of perturbative
QCD (pQCD) becomes possible. The initial particle production is
expected to be dominated by minijets, i.e. partons with $p_T\sim
1\dots 2 $GeV$ \gg \Lambda_{QCD}$ \cite{BM87}.  By assuming
independent multiple semi-hard parton-parton collisions the average
energy carried by minijets with $p_T\ge p_0$ at the rapidity interval
$\Delta Y$ in a central ${\bf b}=0$ $AA$ collision is given in leading-order 
(LO) by \cite{EKL89}

\begin{eqnarray}
\nonumber
\overline{E}_{AA}^T&=&T_{AA}({\bf 0})\sigma\langle
 E_T\rangle\\
&=&T_{AA}({\bf 0})\sum_{\scriptsize{q,\bar{q},g}}
\int_{\scriptsize p_0,\Delta Y}dp_tdy_1dy_2
x_1f_{i/p}(x_1,Q^2)x_2f_{j/p}(x_2,Q^2)\frac{d\hat{\sigma}^{ij\rightarrow jk}}
{d\hat{t}}p_T,
\label{e_t}
\end{eqnarray}

where $T_{AA}({\bf b})$ is the standard nuclear overlap function and
$p_0$ is the smallest transverse momentum scale to be
considered. Collinear factorization is assumed to hold and any effects
beyond it are neglected.  This perturbative minijet approach suffers
from two major sources of uncertainties: \\ ({\it i}) The
next-to-leading order (NLO) corrections, which have not been known
prior to \cite{LO89,ET00}, but rather have been simulated by {\it ad
hoc} $K$-factors.\\ ({\it ii}) The determination of which value of
$p_0$ to use at, say, RHIC or LHC energies. Should one stick to some
constant universal value of $p_0$ or will this parameter possess some
nontrivial $\sqrt s$- and $A$-dependence?

In the following we will provide answers to both of these questions
and combine them to obtain numerical estimates of average transverse
energies and charged particle multiplicities at RHIC and LHC energies.

As is evident from formula (\ref{e_t}), to deal with the first
uncertainty, the relevant quantity to compute in NLO pQCD
is the $\sigma\langle E_T\rangle$, the first moment of the
perturbative $E_T$ distribution in a pp-collision. The infrared safe
NLO computation of this quantity has been presented in \cite{ET00},
where the computation was formulated via the subtraction algorithm of
S. Ellis, Z. Kunszt and D. Soper \cite{KS92}.

\begin{figure}[h!tb]
\vspace{-1.0cm}
\centering
\includegraphics[width=6.2cm,trim=0 -15 0 0]
{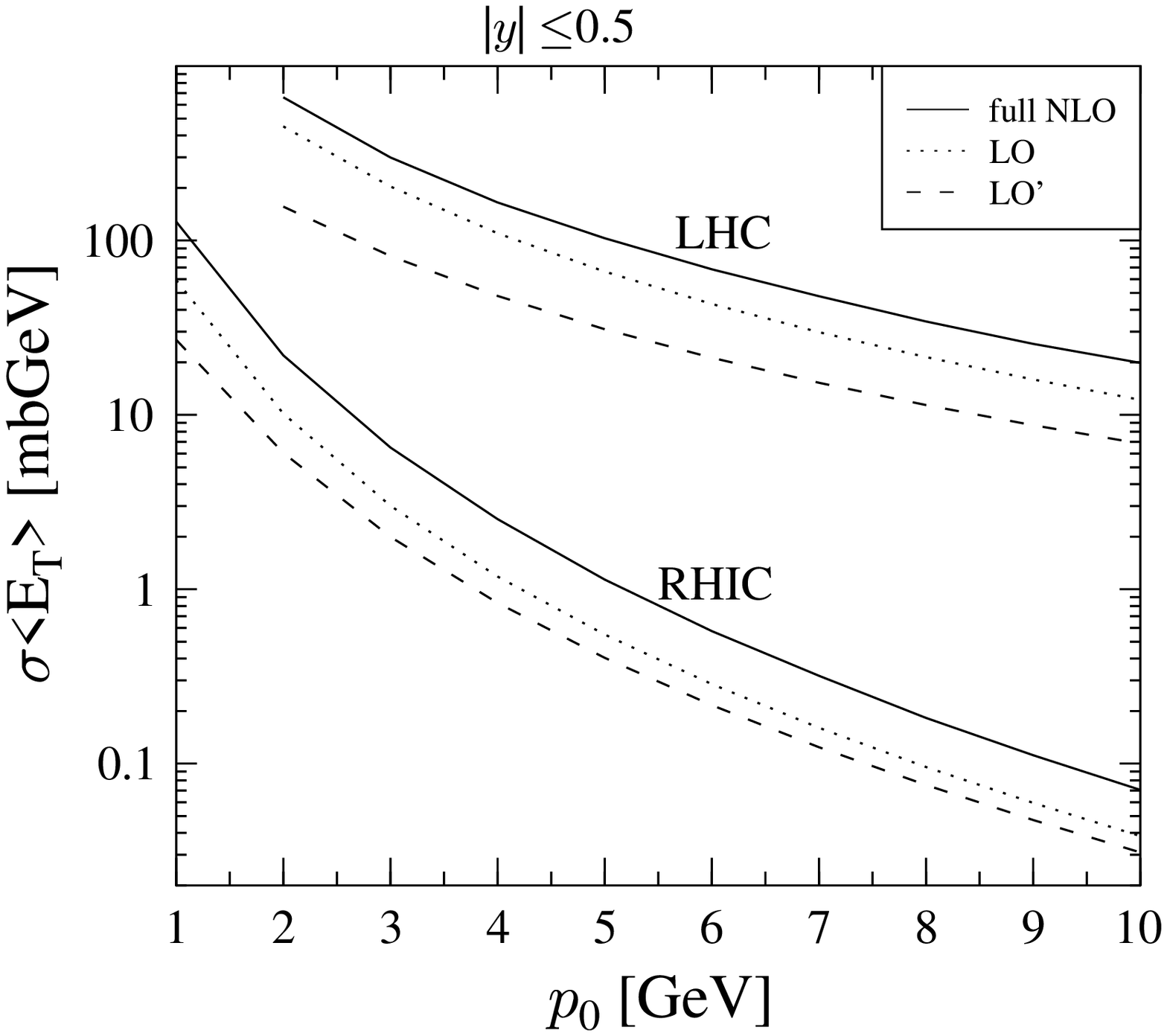}
\includegraphics[width=6.7cm]{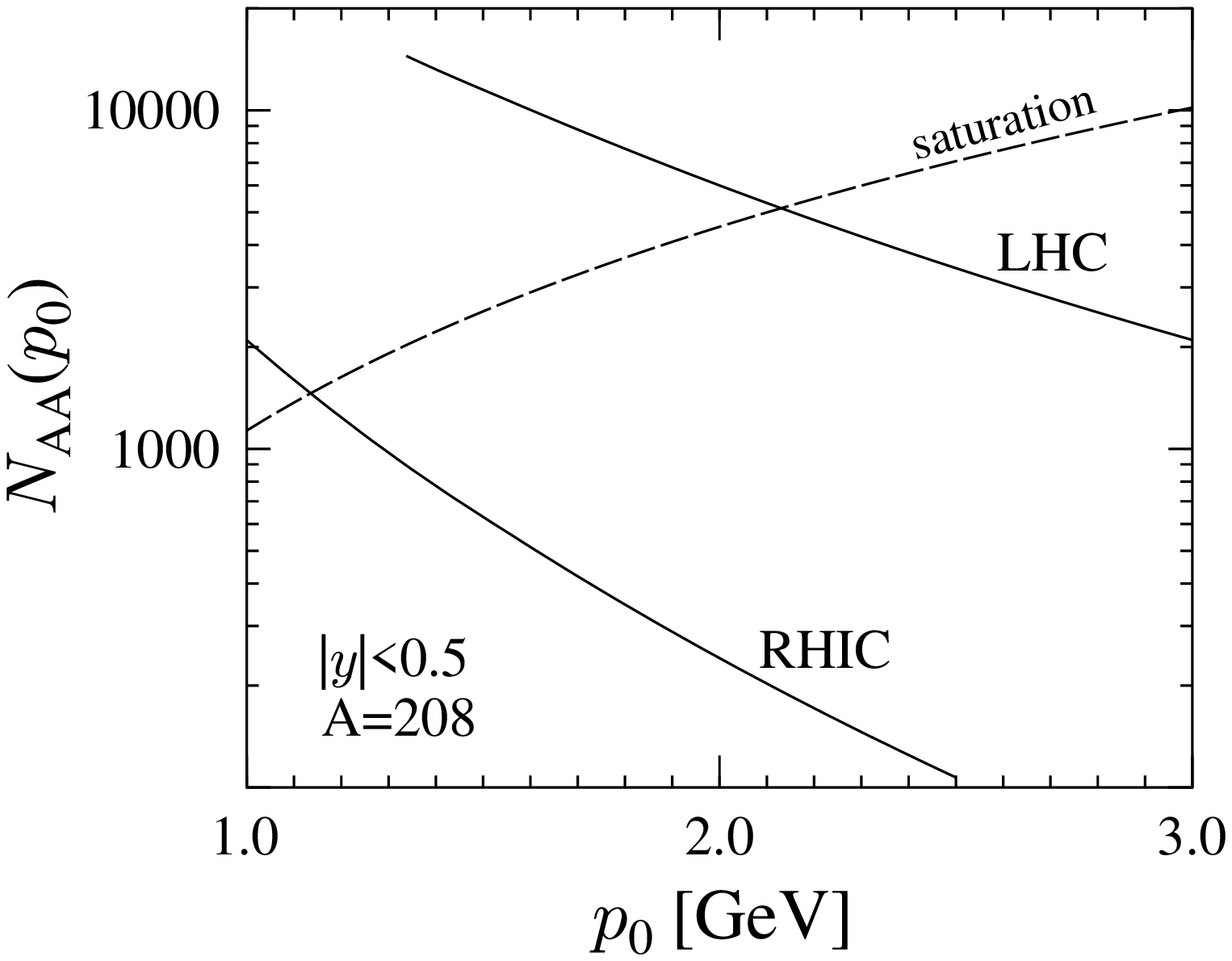}

\vspace{-1.0cm}
\caption[a]{\small (a) The NLO $\sigma\langle
	E_T\rangle$. The LO$^{\prime}$ stands for the leading order result
	evaluated with 2-loop $\alpha_s$ and NLO parton distributions,
	whereas the LO stands for the leading order result evaluated
	with 1-loop $\alpha_s$ and LO parton distributions. The
	rapidity interval is chosen to be the central unit, and the
	parton distributions used are those of GRV-94 -set
	\cite{GRV94}. (b) The average number of QCD-quanta produced
	with $p_T\ge p_0$ and $|y|\le 0.5$ as a function of $p_0$. The
	saturation scale $p_{\rm{sat}}$ is determined by the points of
	intersection with the dashed curve $(p_0^2 R_A^2)$ labelled
	``saturation''.  }
\label{fig1}
\end{figure}

The $E_T$ in central rapidity region is defined to be the total $p_T$
entering this region and originating from hard subprocessess in which
at least an amount of $2p_0$ of transverse momentum is released. The
numerical results are shown in fig. \ref{fig1}(a).
\vspace{0.5cm}

To analyze the implications of these numbers, let us define two
$K$-factors: $K=\textrm{(full NLO)}/\textrm{LO}$ and
$K^{\prime}=\textrm{(full NLO)}/\textrm{LO}^{\prime}$, the first of
these measuring the deviation of the NLO results from the consistent
LO calculation and the latter measuring rather the relative difference
between two subsequent terms of the perturbation series. The magnitude
of these $K$-factors is due to a new kinematical region which
manifests itself only at NLO. If one rejects this new domain, the
resulting $K$-factors would be a factor of two smaller than those
obtained from fig. \ref{fig1}(a), but then a significant amount of
perturbatively calculable $E_T$ would be neglected.

For a detailed description of the calculation and issuses such as the
scale choices, see ref. \cite{ET00} and references therein. For the
purposes to be considered here, the sufficient observation is that we
now have control over the magnitude of NLO corrections which are
stable relative to LO results even at few GeV scales, thus signalling
the applicability of pQCD in this domain.

Turning to the uncertainty ({\it ii}), then, it is clear that as $p_0$ is
decreased the cross section, as well as the uncertainty, grows. At
certain value of saturation, $p_0=p_{\rm{sat}}$ the system becomes very
dense and new physics enters \cite{GRL83}. For large nuclei and
large collision energies this may happen already in the perturbative
domain, which we concluded on the basis of the NLO analysis to include
also the few GeV region. Then the corresponding values for the number of
particles as well as for the amount of $E_T$ are easily produced via
the perturbative computation at $p_0=p_{\rm{sat}}$, which effectively
accounts for the contributions of all scales, since the partons with
$p_T\gg p_{\rm{sat}}$ are rare and those with $p_T\ll p_{\rm{sat}}$, altough
numerous, contribute negligibly to total $E_T$. 

Various ways to determine the actual magnitude of $p_{\rm{sat}}$ can
be conjectured. A simple geometric criterion has been presented in
\cite{EKRT99}. This is based on the idea that if one assigns an
effective area $\pi/p_0^2$ to each gluon produced, then at certain
value of $p_0$ the total area of $N_{AA}(p_0,\sqrt s \Delta Y)$ gluons
produced will exceed the effective transverse area $\pi R_A^2$ of the
nucleus. Therefore one can iterate the equation $N_{AA}=p_{\rm{sat}}^2
R_A^2$ to determine $p_{\rm{sat}}$ for given $A$ and $\sqrt s$, see
fig.\ref{fig1}(b). On the basis of the NLO analysis, we take here $K=2$
to account for the NLO corrections and also implement nuclear
shadowing via the EKS98 parametrization \cite{EKS98}.

All the initial quantities then computed can well be fitted by a
scaling law of a type $CA^b(\sqrt s)^b$. In particular one finds that
\begin{eqnarray}
p_{\rm{sat}}/{\rm GeV} &=& 0.208A^{0.128}(\sqrt s)^{0.191} \nonumber \\ 
\epsilon_i/{\rm (GeV/fm^3)}&=& 0.103A^{0.504}(\sqrt s)^{0.786} 
\label{scalings} \\  
n_i\cdot {\rm fm^3} &=& 0.370A^{0.383} (\sqrt s)^{0.574} \nonumber
\end{eqnarray}
where the particle and energy densities are evaluated at $p_{\rm{sat}}$,
and the whole production process is then considered to take place at
$\tau_0=1/p_{\rm{sat}}$. 

These, however, are just the initial numbers at 0.2 (0.1) fm/c at RHIC
(LHC), and the major problem is how to get from these to the ones at
later instants and to finally arrive at experimentally visible
quantities. Let us therefore assume, not completely without reason
(see \cite{EKRT99}), that the system is initially thermalized in the
sense that it possesses a correct ratio of energy per particle as far
as the dominant gluonic particle content is considered, and expands
conserving the total entropy $S\approx 3.6 N_i$. As the final
particles consist dominantly of pions, for which $S\approx 4N_f$, we
find that $N_f=0.9N_i$. As the initial volume is $V_i=\pi R_A^2 \Delta
 Y/p_{\rm{sat}}$, formulae (\ref{scalings}) give 

\begin{equation}
N_f=1.245 A^{0.92}(\sqrt s)^{0.383}. 
\label{nf}
\end{equation}

After the conference the very first measurements by PHOBOS
collaboration have been announced \cite{PHOBOS}. According to them,
the charged particle multiplicity at midrapidity is $dN/d\eta=408\pm
12\pm 30$ at 56 AGeV and $555\pm 12\pm 35$ at 130 AGeV. The framework
described here gives $N_{ch}=2/3 N_f=370$ and 530 per unit $\eta$
respecti\-vely, when taking into account that number of participants
was reported to be 330 for $\sqrt s=56$ AGeV and 343 for $\sqrt s=130$
AGeV, and that $dN/dy=1.15 dN/d\eta$.

The final $E_T$ in this scenario is obtained by means of
hydrodynamics \cite{EKRT99,KRMcLG} as 
\begin{equation}
E_T=N_f\times[0.39+0.061\ln(N_f/A)]. 
\end{equation}

Numerical values then per unit $\eta$ are $E_T=260$ GeV for $\sqrt
s=56$ AGeV and $E_T=390$ GeV for $\sqrt s=130$ AGeV using the
multiplicites computed above and the quoted participant
numbers. Measurements of these final transverse energies are awaited
to appear soon. These measurements will then allow us to draw conclusions
on the issues such as the true degree of thermalization in the system.  

\vspace{0.5cm}

\noindent {\bf Acknowledgements:} I wish to thank K.J. Eskola, K. Kajantie and
P.V. Ruuskanen for collaboration. The financial support from the
Academy of Finland as well as from the organizers of the CIPANP2000
conference is gratefully acknowledged.

\end{document}